\documentclass[12pt,preprint]{aastex}
\usepackage{xspace}
\usepackage{graphicx}
\usepackage{natbib}
\usepackage{url}
\newcommand{\hide}[1]{}

\newcommand{\expo}[1]{\ensuremath{10^{#1}}\xspace}

\newcommand{\gsim}{\ensuremath{\,\gtrsim\,}\xspace}
\newcommand{\lsim}{\ensuremath{\,\lesssim\,}\xspace}
\let\phs\undefined
\newcommand\phs{\phantom{$-$}}%
\newcommand{\gl}{\ensuremath{\ell}\xspace}
\newcommand{\gb}{\ensuremath{{\it b}}\xspace}
\newcommand{\absb}{\ensuremath{\vert\,\gb\,\vert}\xspace}
\newcommand{\vlsr}{\ensuremath{V_{\rm LSR}}\xspace}
\newcommand{\lb}{\ensuremath{(\gl,\gb)}\xspace}
\newcommand{\lv}{\ensuremath{(\gl,v)}\xspace}

\newcommand{\dec}{\ensuremath{\delta}\xspace}

\newcommand{\kms}{\ensuremath{\,{\rm km\,s^{-1}}}\xspace}

\newcommand{\cm}{\ensuremath{\,{\rm cm}}\xspace}
\newcommand{\pc}{\ensuremath{\,{\rm pc}}\xspace}
\newcommand{\kpc}{\ensuremath{\,{\rm kpc}}\xspace}

\newcommand{\mK}{\ensuremath{\,{\rm mK}}\xspace}

\newcommand{\ghz}{\ensuremath{\,{\rm GHz}}\xspace}

\newcommand{\degree}{\ensuremath{\,^\circ}\xspace}

\newcommand{\jy}{\ensuremath{\,{\rm Jy}}\xspace}

\newcommand{\mjy}{\ensuremath{\,{\rm mJy}}\xspace}

\newcommand{\hi}{{\rm H\,{\scriptsize I}}\xspace}
\newcommand{\hii}{{\rm H\,{\scriptsize II}}\xspace}

\newcommand{\hal}[1]{\ensuremath{{\rm H}\,{#1}\,\alpha\xspace}}
\newcommand{\hna}{\ensuremath{{\rm H}\,{\rm n}\,\alpha\xspace}}
\newcommand{\hnaa}{\ensuremath{\langle\,\hna\,\rangle\xspace}}

\newcommand{\cor}{\ensuremath{^{\rm 13}{\rm CO}}\xspace}

\newcommand{\hiea}{{\rm H\,}{{\scriptsize I}{\rm~E/A}}}

\shorttitle{Arecibo \hii\ Region Discovery Survey}
\shortauthors{Bania et al.}

\begin{document}

\title{THE ARECIBO H\,{\small \bf II} REGION DISCOVERY SURVEY}

\author{T.~M.~Bania\altaffilmark{1}, L.~D.~Anderson\altaffilmark{2}, 
and Dana~S.~Balser\altaffilmark{3}}
\altaffiltext{1}{Institute for Astrophysical Research, Department of
  Astronomy, Boston University, 725 Commonwealth Ave., Boston MA
  02215, USA}
\altaffiltext{2}{Department of Physics, West Virginia University,
  Morgantown, WV 26506, USA}
\altaffiltext{3}{National Radio Astronomy Observatory, 520 Edgemont
  Road, Charlottesville VA, 22903-2475, USA}


\begin{abstract}

We report the detection of radio recombination line emission (RRL)
using the Arecibo Observatory at X-band (9\,GHz, 3\,cm) from 37
previously unknown \hii\ regions in the Galactic zone $66\degree \ge
\gl \ge 31\degree$ and $\absb \le 1\degree$.  This Arecibo
\hii\ Region Discovery Survey (Arecibo HRDS) is a continuation of the
Green Bank Telescope (GBT) HRDS.  The targets for the Arecibo
HRDS have spatially coincident 24\,\micron\ and 20\,cm emission of a
similar angular morphology and extent.  To take advantage of Arecibo's
sensitivity and small beam size, sources in this sample are fainter,
smaller in angle, or in more crowded fields compared to those of the
GBT HRDS.  These Arecibo nebulae are some of the faintest \hii regions
ever detected in RRL emission.  Our detection rate is 58\%, which is
low compared to the 95\% detection rate for GBT HRDS targets.  We derive
kinematic distances to 23 of the Arecibo HRDS detections.  Four
nebulae have negative LSR velocities and are thus unambiguously in the
outer Galaxy.  The remaining sources are at the tangent point distance 
or farther. 
We identify a large, diffuse \hii\ region complex that has an
associated \hi\ and \cor shell.  The $\sim\,$90\,\pc diameter
of the G52L nebula in this complex may be the largest Galactic
\hii\ region known, and yet it has escaped previous detection.  
%
%
\end{abstract}

\section{Introduction}
\hii\ regions are the zones of plasma surrounding massive OB stars.
Because such stars only live for a few Myr, \hii\ regions are clear
indicators of ongoing massive star formation.  They are luminous
objects at radio and infrared (IR) wavelengths and can be seen across
the entire Galactic disk.  Despite much effort, the census of
Galactic \hii\ regions is still incomplete.  A complete census of
Galactic \hii\ regions is important if we are to use these nebulae to
trace Galactic structure, to determine the Galactic high-mass
star formation rate, and to characterize the Galactic
\hii\ region population as a whole.  

Radio emission from \hii\ regions is dominated by two emission mechanisms.  
\hii\ region plasma emits radio continuum emission from thermal
Bremsstrahlung.  Recombining electrons and nuclei in the plasma
produce recombination line emission, which is observable from optical
to radio wavelengths.  Most surveys of \hii\ regions over the past few
decades were conducted using radio recombination line (RRL)
emission, and the targets were identified from radio continuum
maps \citep[e.g.,][]{reifenstein70, wilson70, caswell87, lockman89}.
Measuring a RRL gives a radial velocity that, when combined with a
model of Galactic rotation, can be used to derive a kinematic
distance \citep[e.g.][]{wilson70}.

Dust in the \hii\ region emits in the IR, especially the mid- to
far-IR.  This was demonstrated conclusively by \citet{wc89a} who
measured radio continuum emission from ultra-compact \hii\ regions
spatially coincident with IRAS point sources.  The combination of
infrared and radio continuum emission strongly indicates that a source
is emitting thermally, i.e., it is an \hii\ region or a planetary
nebula (PN) and not a supernova remnant or an active galactic nucleus
\citep{haslam87, broadbent89}.
More recent RRL surveys of \hii\ regions used IR data to identify 
candidate targets \citep{araya02,  watson03, sewilo04, anderson11}.

The \hii\ Region Discovery Survey (HRDS) is an ongoing census program
that aims to locate and measure the RRL emission from as many Galactic
\hii\ regions as possible.  In the Green Bank Telescope (GBT) HRDS, we
detected the RRL and continuum emission from 448 previously unknown
Galactic \hii\ regions at X-band (9\,GHz, 3\,cm).  This census doubles
the number of known \hii\ regions over the Galactic zone studied.
\citet[][hereafter Paper~I]{bania10} described the GBT HRDS survey.
\citet[][hereafter Paper~II]{anderson11} gave the GBT HRDS methodology
and provided the HRDS source catalog.  \citet[][hereafter
  Paper~III]{anderson12c} derived kinematic distances to a large
fraction of the GBT HRDS sample.  T. Wenger, et al. (2012, in prep.),
will give the properties of the helium and carbon RRLs detected in the
GBT HRDS nebulae.

 Here we report the discovery of 37 previously unknown \hii\ regions
found by detecting RRL emission from targets using the Arecibo
telescope at X-band (9\ghz, 3\cm).  With $\sim9$ times the collecting
area of the GBT, Arecibo has a higher sensitivity.  Its beam is nearly
three times smaller than that of the GBT.  These factors make it
possible for Arecibo to find \hii\ regions that are fainter and
smaller than GBT HRDS targets or are located in confused regions of
the Galaxy.

\section{Target Sample}
We define the Arecibo target sample in the same way as for the
GBT HRDS, searching by eye for spatial coincidences between
mid-infrared (MIR) and radio continuum sources.  Sources with
spatially coincident MIR and radio continuum emission are likely to be
thermal emitters; the 95\% detection rate of the GBT HRDS targets so
defined confirms this.  We use {\it Spitzer} MIPSGAL data
\citep{carey09} for the MIR and both 20\,cm MAGPIS \citep{helfand06}
and 21\,cm VLA Galactic Plane Survey data \citep[VGPS;][]{stil06} for
the radio.  MAGPIS has an angular resolution of $\sim5\,\arcsec$,
comparable to the MIPSGAL resolution of $6\,\arcsec$, whereas the VGPS
has an angular resolution of $1\,\arcmin$.  Because the Arecibo beam at
X-band, $\sim30\,\arcsec$, is smaller than the resolution of the VGPS
data, MAGPIS is helpful when attempting to identify candidates in
crowded fields of IR and radio continuum emission.  We show in
Figure~\ref{fig:three_color} three-color {\it Spitzer} MIR images for
all the new Arecibo \hii\ region detections.

The Arecibo targets are fainter, smaller, or lie in more crowded fields
compared with the GBT HRDS sources. The Arecibo gain at X-band is
$\sim$\,2.5 times that of the GBT; Arecibo can therefore in principle
detect individual spectral lines from fainter sources than the GBT.
The Arecibo beam at X-band is $\sim30\,\arcsec$ compared to
$\sim82\,\arcsec$ for the GBT.  For sources compact with respect to the
GBT beam, Arecibo will suffer less beam dilution and therefore should
be more sensitive.  In complicated fields, the GBT beam may include
the emission from multiple \hii\ regions along the line of sight.  For
these fields, the smaller Arecibo beam size can better measure the
emission from a single \hii\ region.

Our \hii\ region candidate targets are located in the Galactic zone
$66\degree \ge \gl \ge 31\degree$ and $\absb \le 1\degree$.  These
limits are set by the sky coverage of the Arecibo telescope and the
MIPSGAL survey boundaries.  The Arecibo declination limit of $\dec >
-1\degree\ 20\,\arcmin$ prevents observations of the Galactic plane below
$\gl \sim 31\degree$.  The upper longitude limit and the latitude
limit are set by the MIPSGAL survey boundaries.  Finally, the MAGPIS 
survey extends only to $\gl < 48\,\fdg5$ so we cannot use these 
20\,\cm data for all candidates.

We observe the 63 targets that have the brightest expected X-band
continuum fluxes.  We compute these fluxes from aperture photometry of
21\,\cm VGPS data. These L-band fluxes are extrapolated to X-band
under the assumption that the emission at 21\,cm is optically thin.
In the GBT HRDS we found that these extrapolated values disagree by as
much as $\sim\,\pm$100\% from the measured GBT X-band continuum
intensities, possibly because the derived fluxes are poor or because
the assumption that all sources are optically thin at 20\,cm is
invalid.

We detect RRL emission from 37 targets.  Table~\ref{tab:sources} lists
these \hii\ regions and gives the source name, its Galactic and
Equatorial (J2000) coordinates, the velocity of detected CS
$2\rightarrow 1$ emission from \citet{bronfman96}, and a
characterization of the source morphology based on {\it Spitzer}
GLIMPSE 8.0\,\micron\ images \citep{benjamin03}.  The morphological
classifications are the same as in Paper~II.  If the source is
associated with a GLIMPSE bubble from \citet{churchwell06}, we give
the bubble name in parentheses in the ``Morphology'' column.  Because
CS emission is only detectable for dense molecular clouds, it is a
good tracer of the early stages of star formation.  Paper~II found
good agreement between the CS and RRL velocities and also gave
associations with 6.7\,GHz methanol masers from \citet{pestalozzi05}.
No Arecibo HRDS sources have a methanol detection within $2\,\arcmin$.

\section{Observations and Data Analysis}
To make the Arecibo HRDS we used the 305\,m telescope during the
period October 2011 to April 2012.  Each observing session consists of
one sidereal pass of the first quadrant Galactic plane and lasts
approximately five hours.  At the beginning of each run we make radio
continuum cross-scans of the standard source B1843+09 to verify the
observing setup, and to confirm that the measured gain is within the
expected range, i.e., that the telescope is in focus.  We also verify
the spectral tunings by observing the bright \hii\ region
G045.455+0.059. As in Paper~II, we use our custom IDL software package
TMBIDL\footnote{See \url{http://www.bu.edu/iar/tmbidl}} to analyze
these data.

The telescope gain depends on what part of the dish is illuminated.
Because we are observing at the high frequency limit of Arecibo, the
changing rms of the surface that is being illuminated as a source is
tracked can alter the gain by 20\% or more.  Furthermore, at low
zenith angles, the telescope cannot track fast enough to remain
pointed at the source, whereas at high zenith angles there is
spillover.  Both these situations also affect the telescope gain.  The
Arecibo gain at X-band is in the range $\sim4-6$\,K$\,$Jy$^{-1}$,
depending on azimuth and zenith angle\footnote{See
  \url{http://www.naic.edu/~astro/RXstatus/Xband/RX\_xband.shtml}}.
This gain variation motivates many of our choices for the
observational setup and subsequent data analysis.  First, we do not
observe radio continuum from our sources as we did in the GBT HRDS.
Such measurements would not be reliable without extensive calibration,
and even then would likely be of low quality given the weather
conditions we typically encounter.  Second, we do not calibrate
measured intensities with astronomical sources of known intensity, but
instead rely solely on noise diodes to measure the system temperature.
This again is difficult to do correctly with Arecibo, and would add
little to the astrophysical importance of these data.  The only
reliably determined line parameter is the RRL LSR velocity.  The RRL
line intensity and line width should be used only with great caution.

We use the ``Interim'' correlator configured into sub-bands that are 
tuned to four consecutive RRL transitions, \hal{89} to \hal{92}.  We
sample both circular polarizations for each tuning.  We thus get 8
independent spectra per observation: 4 RRLs $\times$ 2 orthogonal
polarizations each.  (For comparison, the GBT HRDS observed seven RRL
transitions simultaneously, \hal{87} to \hal{93}.)  Each sub-band has
a 25\,MHz total bandwidth and its 1024 channels are 9-level sampled.
The velocity resolution of all sub-bands is $\sim\,0.8\,\kms$ per
channel.  Because the source velocities are unknown, we tune all
sub-bands to a 0\,\kms\ LSR center velocity.  which gives us a velocity
coverage of $\sim\,\pm$ 400\,\kms.  Over the survey zone, this
velocity range is more than sufficient to detect emission from all
Galactic \hii\ regions bound to the Galaxy.

We observe in total-power, position switching mode with On- and
Off-source integrations of five minutes each, hereafter a ``pair.''
The Off-source position is chosen so that we track the same azimuth
and zenith angle path as the On-source scan does.  We set the total
integration time individually for each source based on the strength
(or absence) of RRL emission.  Total source integration times
range from 1200 to 5400 sec, with an average of 1800 sec.  Each
individual pair is calibrated by firing the noise diodes.

  About 2 hours of data were
unusable due to rain storms and it was cloudy or lightly raining for
most observing runs.  X-band is the highest radio frequency band
observable with Arecibo and focusing the telescope is especially
important compared to the lower frequencies.  The Arecibo telescope is
focused by using ``tie down'' cables running from each vertex of the
triangular receiver platform to the ground.  For 20\% of our observing
runs, the tie down system was not operational and the focus was not
optimal.  Finally, two of our observing runs are plagued by baseline
ripples.  These ripples appear in all sub-bands, although they are
only readily apparent in the right circularly polarized data.  The
ripples have amplitudes of $\sim 5$\,mK and are present for most (but
not all) of the two aforementioned days.  We expunge almost all of
these data in subsequent analyses.  We do not know the source of these
ripples.

To first order, RRLs at X-band all have the same intensity, line
width, and velocity; they can therefore be combined to create more
sensitive \hnaa\ spectra.  Because of their tunings, each sub-band has
a slightly different spectral resolution, which complicates the
process of averaging the individual spectral lines.  In Paper~II, we
followed \citet{balser06} and interpolated data from all sub-bands to
that of the sub-band with the lowest velocity resolution.  Here, for
the \hal{89} sub-band the spectral resolution is 0.798\,\kms, whereas
it is 0.881\,\kms\ for the \hal{92} sub-band.  For an average RRL line
width of 25\,\kms\ (see Paper~II), the difference between the \hal{89}
sub-band and the \hal{92} sub-band for this line width is 3 channels,
or $\sim2.5$\,\kms.  Because we smooth the \hnaa\ spectra to a
4.2\,\kms\ resolution (see below), the difference in velocity
resolution between the four sub-bands is unimportant and we do not
perform any interpolation.  All four sub-bands are Doppler tracked and
have the same center velocity.  The line velocities are therefore
unaffected by the different spectral resolutions.  We calculate
average spectra in the standard way with a weighting factor of $t_{\rm
  intg}/T_{\rm sys}^2$, where $t_{\rm intg}$ is the integration time
and $T_{\rm sys}$ is the total system temperature.
Figure~\ref{fig:four_ifs} shows an example of our analysis for the
\hii\ region G034.132+0.472.  The black lines show spectra for the
four tunings (sub-bands).  Each is the average of both circular
polarizations.  The average \hnaa\ spectrum is drawn in red.

We smooth the \hnaa\ spectra using a normalized Gaussian function of
FWHM five channels to give a spectrum with a velocity resolution of
4.2\,\kms.  RRLs average $\sim25$\,\kms\ FWHM (see Paper~II) so this
guarantees that for most sources we have at least five spectral
resolution elements across the line.  We remove a polynomial baseline
from the combined, smoothed spectrum.  We determine the polynomial
baseline order individually for each source, but it is generally third
order or less.  We then fit Gaussian functions to detected lines
individually for each source.  This gives us the LSR velocity, line
intensity, and line width for each emission component.  There are no
other strong atomic, molecular, or unidentified lines known in these
frequency bands and therefore the brightest detected lines are
hydrogen RRLs.
Finally, the Arecibo beam size changes with frequency so we might be
sampling slightly different volumes of space with each tuning.
Between the four sub-bands, the Arecibo beam is 10\% different.  As in
the GBT HRDS, we do not attempt to correct for this effect.

\section{The Arecibo H\,{\bf\footnotesize II} Region Discovery Survey Catalog
\label{sec:catalog}}
We detect RRL emission from 37 sources out of 64 observed targets, a
58\% success rate.  We show these \hnaa\ spectra in
Figure~\ref{fig:spectra}.  Two sources have two RRL velocity
components so the Arecibo HRDS contains 39 discrete RRLs altogether.
The average rms noise for the survey \hnaa\ spectra is 2.0\,mK.  Given
the Arecibo X-band gain of $\sim4$\,K\,\jy, this corresponds to
0.5\,\mjy.  For comparison, the GBT HRDS has a 95\% detection
rate and an average rms of 1.0\,\mjy.

Table~\ref{tab:line} gives the Arecibo HRDS catalog hydrogen RRL line
parameters, which come from Gaussian fits to the \hnaa\ spectra.  It
lists the source name, Galactic co-ordinates, line intensity, FWHM
line width, LSR velocity, and the rms noise in the spectrum.  Errors
in line parameters are the $\pm~1~\sigma$ uncertainties from the Gaussian
fits.  As we describe in Section 3, only the LSR velocity is reliably
determined.  The errors given in Table~\ref{tab:line} for the line
intensity and FWHM line width underestimate the true errors in these
parameters.  For the two sources with multiple RRL velocity
components, we append an ``a'' or a ``b'' to the source names, in
order of decreasing LSR velocity.  We list the sources for which we
did not detect RRL emission in Table~\ref{tab:nondetections}, which
gives the source name, Galactic position, rms noise, and GLIMPSE
8\micron\ morphology.


We re-observed a sample of GBT HRDS nebulae in order to verify the
telescope configuration, to evaluate the data processing accuracy, and
to cross-calibrate the two surveys.  Table~\ref{tab:comparison} lists
hydrogen RRL parameters for four nebulae common to both surveys.
Because of the different beam sizes, observations with the GBT and
Arecibo potentially sample different volumes for a given source so we
do not necessarily expect the line parameters to agree exactly.  We
are also comparing antenna temperature and not flux density, so we
would not expect the line intensities to agree in these units.
Moreover, given the gain variations of the Arecibo telescope,
comparing the line intensities provides no useful insights. The
agreement is, however, quite good for the RRL velocity and line width,
although it is best for the brightest nebula, G034.133+0.471.  For
these sources, the mean absolute difference in velocity is 0.6\,\kms
and the mean absolute difference in FWHM line width is 2.7\,\kms.  The
line width is larger for all the Arecibo observations, which is due to
how the four RRLs are averaged.

%


\section{Distances}
We derive kinematic distances to the Arecibo HRDS nebulae as in Paper~III.
Kinematic distances are computed from models of Galactic
rotation that take as inputs the source Galactic longitude and LSR
velocity.  Although some models also use the Galactic latitude to
account for vertical deviations in Galactic rotational velocities,
\citep[e.g.,][]{levine08}, we do not use them.  As was also the case
for the GBT HRDS, kinematic distances are the only method that can be
used to derive distances to the majority of Arecibo HRDS sources.
We compute all kinematic distances here using the \citet[hereafter
B86]{brand86} rotation curve.

For sources in the inner Galaxy, there are two possible kinematic
distances for each measured velocity, a ``near'' and a ``far''
distance.  The ``near'' and the ``far'' distance for a given line of
sight are spaced equally about the tangent-point distance, which is
the location of maximum radial velocity.  This problem is known as the
kinematic distance ambiguity, KDA.  Without additional information, it
is unknown which distance is correct for a given nebula.  The outer
Galaxy has no KDA, however, so our first Galactic quadrant Arecibo
HRDS sources with negative RRL velocities are unambiguously in the
outer Galaxy.  We can derive kinematic distances for them directly
from their measured velocities.

The \hi\ Emission/Absorption (\hiea) method has proved to be very
effective at resolving the KDA for \hii\ regions
\citep[see][]{kuchar94, kolpak03, anderson09a}.  In Paper~III we
determined distances to 158 GBT HRDS sources by the \hiea\ method
using \hi\ data from the VGPS.  The \hiea\ method relies on the
detection of \hi\ absorption of the broadband thermal radio continuum
emission from an \hii\ region.  \hi\ is ubiquitously distributed in
the Galaxy and so it emits at all velocities permitted by Galactic
rotation.  If \hi\ absorption is found between the nebular RRL
velocity and the tangent point velocity, this favors the far distance.
The absence of any \hi\ absorption beyond the nebular RRL velocity
favors the near distance.  For the \hiea\ method to be successful,
however, the \hii\ region must be bright enough to allow for
detectable absorption.

We are able to derive kinematic distances for 23 Arecibo
\hii\ regions, which is 62\% of the HRDS sample.  Only three, however,
have bright enough continuum emission to make an \hiea\ analysis.  For
the other sources, we can establish kinematic distances to them
because they are near the tangent-point distance, in the outer Galaxy
(or very near to the Solar orbit), or associated with large
star-forming complexes with previously known distances.

Two of the \hiea\ sources, G052.201+0.752 and G052.259+0.700, are in
the G52 complex (see Section~\ref{sec:G52}).  Both their RRL
velocities are near 10\,\kms\ and \hi\ absorption is detected up to
60\,\kms, thus favoring the far distance. The third source,
G041.881+0.493, has a RRL velocity of 23.2\,\kms\ and \hi\ absorption
is detected up to 68\,\kms.  This again favors the far distance.
Paper~III assigned ``quality factors'' (QFs) to assess the reliability
of an \hiea\ KDA determination. Using these same qualitative criteria,
the \hiea\ KDA resolution for these three nebulae would all be of the
highest-quality, QF ``A''.

As we did in \citet{anderson09a} and Paper~III, we assign the
tangent-point distance to the 12 Arecibo HRDS sources whose velocities
are within 10\,\kms\ of the tangent-point velocity.  For sources near
the tangent-point, our ability to discriminate between the near and
far kinematic distances using the \hiea\ method is limited.
Furthermore, as the LSR velocity approaches the tangent-point
velocity, the difference between the near and far distances becomes
small so resolving the KDA is less important.

The four Arecibo HRDS sources with negative RRL velocities are
unambiguously in the outer Galaxy.  One of these, G034.591+0.244, has
two RRL velocity components: 56.6\,\kms\ and $-19.4\,\kms$.  We
believe that the latter is the actual velocity of this nebula and that
the 56.6\,\kms component comes from diffuse gas along the line of
sight.  This is because there is no evidence for any spatially
extended, diffuse gas in the outer disk for this part of the Milky Way
whereas the GBT HRDS catalog shows many examples of extended, diffuse
gas in the first quadrant inner Galaxy. This negative velocity for
G034.591+0.244 then gives a kinematic distance of 16.1\,kpc.

There are three nebulae in the G46 complex (see
Section~\ref{sec:individual}) that have very low RRL velocities, $\vlsr
\lsim 10$\,\kms.  Such low velocities give near kinematic distances
very close to the Sun, \lsim 0.5\,kpc.  Given their low fluxes and
small angular sizes, however, all these sources are likely to be at
their far kinematic distances.
Finally, G037.468$-$0.105 is spatially coincident with the W47
\hii\ region complex, and its RRL velocity agrees with other nebulae
in the complex.  \citet{anderson09a} derived a far kinematic distance
of 9.6\,kpc\ for these objects so we place G037.468$-$0.105 there as
well.

We give the kinematic distances for the Arecibo HRDS nebulae in
Table~\ref{tab:dist}, which lists for each source its name, LSR
velocity, tangent-point velocity, near and far distances, KDA
resolution, heliocentric distance and its $\pm1\sigma$ uncertainty,
Galactocentric radius, and height above the Galactic plane.  All
kinematic distances and tangent-point velocities are based on the B86
rotation curve.  The distance uncertainties are taken from Paper~III,
which provided percentage uncertainties in kinematic distances for a
range of \lv-space loci.

Table~\ref{tab:dist} shows that of the 19 inner Galaxy nebulae with
kinematic distances, 7 (37\%) are at the far distance, 12 (63\%) are
at the tangent-point distance, and none are at the near distance.  In
comparison, Paper~III found that for GBT HRDS sources 61\% are at the
far distance, 31\% at the tangent point distance, and 7\% at the near
distance.  Over the longitude range of the Arecibo HRDS, these numbers
change to 69\%, 30\%, and 1\%, respectively.  Thus, somewhat
surprisingly since Arecibo is a more sensitive telescope, a smaller
fraction of the Arecibo sample is at the far distance compared to the
GBT HRDS.  The small number statistics of the Arecibo HRDS distance
sample prevent us from drawing any strong conclusions about this.

The dearth of sources at the near distance does, however, imply that
the Arecibo HRDS nebulae are on the whole quite distant.  The average
distance of the Arecibo nebulae is 9.0\,kpc, whereas it is 10.1\,kpc
for the GBT HRDS (9.8\,kpc over the Arecibo longitude range) and
8.4\,kpc for the \citet{anderson09a} \hii\ region sample (8.0\,kpc
over the Arecibo longitude range).  That the GBT HRDS sources are on
average more distant is surprising given the weak line intensities of
the Arecibo HRDS sample (see Section~\ref{sec:intensities}).  Even if
all the Arecibo HRDS sources without derived kinematic distances were
at the far distance, however, the average distance would still be just
9.3\,kpc.

\section{Discussion}
\subsection{Hydrogen Recombination Line Intensities\label{sec:intensities}}
The Arecibo HRDS nebulae are some of the faintest \hii\ regions ever
observed in RRL emission.  The average hydrogen recombination line
intensity is 10.1\mK, or 2.5\mjy assuming a gain of 4\,K\,Jy$^{-1}$.
These are much weaker recombination lines than those in the GBT HRDS,
which has an average line intensity of 35.4\mK, or 17.7\mjy given the
GBT gain of 2\,K\,Jy$^{-1}$ \citep{ghigo01}.  Furthermore, in contrast
to the GBT HRDS, the Arecibo HRDS has {\it only} faint targets.  More
than 74\% of the detected lines have intensities $\lsim10$\,\mK
whereas only 18\% of GBT HRDS sources meet this criterion.  RRL
surveys prior to the GBT HRDS observed even brighter targets with
higher line intensities (see Paper~II, their Figure~9 comparison with
\citealt{lockman89}).  The relatively low detection rate of Arecibo
candidate \hii\ region targets is a direct result of the low mean flux
of the Arecibo HRDS nebulae.

\subsection{Galactic Distribution}
The \lv-distribution of Galactic \hii\ regions projected onto that
part of the first quadrant Galactic plane accessible to the Arecibo
telescope is shown in Figure~\ref{fig:lv}.
Seven Arecibo nebulae have velocities beyond the B86 rotation curve
tangent-point velocity.  They are located in the zone $44\degree \le
\gl \le 60\degree$, and five of them lie in the range $49\,\fdg7 \le
\gl \le 52\,\fdg4$.  The largest discrepancies are for
G051.402$-$0.890, G052.397$-$0.580, and G059.603+0.911, which are
23.7, 20.3, and 15.1\,\kms\ beyond the B86 tangent-point velocity,
respectively.  The location of these five sources in \lv-space is
close to the Sagittarius Arm tangent point.  Similar, but smaller,
velocity deviations beyond the nominal tangent-point velocity can be
seen in the \hi\ emission from this part of the Galaxy \citep[see]
[their feature ``C'']{burton66}.

The Figure~\ref{fig:lv} nebulae are the census of {\em all} currently
known \hii\ regions in this zone. Altogether, in the longitude range
$\gl\,\sim\,50\degree\, - \,52\degree$ there are 18 nebulae with
velocities that are beyond the terminal velocities predicted by all
three of the rotation curve models we use here.  Five \hii\ regions
have velocities well beyond, $\gsim 20 - 25 \kms$, their tangent point
velocities.  These excursions are $\sim50\%$ higher than the models'
average terminal velocity in this direction. This is evidence for
strong streaming motions at this location.

We compare in Figure~\ref{fig:rgal} the Galactic radial distribution
of the Arecibo HRDS nebulae with that of all previously known
\hii\ regions, a sample that includes the GBT HRDS.  Over the Arecibo
longitude range, we compute the Galactocentric radii for both
\hii\ region samples using the B86 rotation curve.  The distribution
of Arecibo sources shows a strong peak near 6\,kpc, similar to what is
seen for the previously known sample.  The Arecibo nebulae also show a
large peak near 8\,kpc that is not seen as strongly for the previously
known sample.  With only 37 sources in the Arecibo HRDS, however,
these findings are not particularly robust.

\subsection{Comments on Individual H\thinspace {\small II} Regions\label{sec:individual}}
{\noindent \bf G037.498+0.530}\\ 
We detect two velocities for this source: 11.6\,\kms\ and 47.1\,\kms.
Both lines are of equal intensity.  This nebula is near the HRDS
source G037.485+0.513, which has three velocities: 4.1\,\kms,
50.0\,\kms, and 90.8\,\kms.  These components are all of comparable
intensity.  \citet{bronfman96} detect one CS component with a velocity
of 10\,\kms from the nearby position \lb\ = (37\,\fdg494, 0\,\fdg530).
This velocity component has also been detected in NH$_3$ by the Red
MSX Survey \citep{urquhart11}.  It therefore seems likely that the RRL
source velocity for G037.498+0.530 is 11.6\,\kms. One possible
scenario is that 90.8\,\kms\ is the source velocity for GBT HRDS
G037.485+0.513 and the $\sim$50\,\kms\ component shared by both
sources is diffuse.  In this scenario, the $\sim$4--12\kms\ emission
is also common to both sources.  Finally, the GBT HRDS also detected
emission from G037.498+0.530.  Further observations are needed to
clarify matters.

{\noindent \bf G043.738+0.114}\\ 
This source has a velocity of 73.1\,\kms\ and is associated with the
GLIMPSE bubble N89.  It is near the HRDS source G043.770+0.070, which
has a velocity of 70.5\,\kms\ and is also associated with a GLIMPSE
bubble, N90 (see Figure~\ref{fig:three_color}).  Neither object has a
distance determination as yet.

{\noindent \bf G046.017+0.264}\\ Because of its MIR morphology and
broad line width, this source might be a planetary nebula.
\citet{urquhart09} classify this source as a PN.  \citet{anderson12a}
show that IR colors can be used to distinguish between \hii\ regions
and PNe.  MIR aperture photometry for this source using GLIMPSE,
MIPSGAL, and WISE \citep{wright10} data, however, is inconclusive.
FIR data would help resolve the nature of this object.

{\noindent \bf G046.176+0.536, G046.203+0.535, \& G046.213+0.548}\\ 
These three sources are in a complex we call here G46 and have
velocities ranging from 2.1\,\kms\ to 6.3\,\kms.  G046.203+0.535 is a
GBT HRDS nebula, but the GBT beam included emission from the other two
sources.  Paper~III did not establish a distance because the
radio continuum emission is too faint.  The small angular size, weak
continuum flux, and low LSR velocities of the nebulae in this complex
favor the far distance.  We thus place all three sources at the far
kinematic distance of 11.5\,kpc.

{\noindent \bf G052.201+0.752 \& G052.259+0.700}\\ 
These two sources and the nearby re-observed GBT HRDS source
G052.160+0.706 all have emission in the off-source direction.  The
source velocities for G052.201+0.752, G052.259+0.700, and
G052.160+0.706 are 8.7\,\kms, 7.4\,\kms, and 8.2\,\kms, respectively.
The off-source velocities are 53.7\,\kms, 51.6\,\kms, and 52.0\,\kms.
We fit the negative-intensity off-source emission simultaneously with
that of the on-source emission.  The fits to the on-source emission
are the line parameters given in Table~\ref{tab:line}.  The lines are
well-separated in velocity (see Fig.~\ref{fig:spectra}) so the line
parameters do not appear to be strongly affected by the off-source
emission.  The derived LSR velocities do not change when the lines are
fit individually.  Furthermore, for G052.160+0.706 the Arecibo and GBT
HRDS velocities match.

The off-source positions, \lb\ = (52\,\fdg884, $-$0\,\fdg505),
(52\,\fdg942, $-$0\,\fdg557), and (52\,\fdg844, $-$0\,\fdg548), 
all coincide with a single large, diffuse \hii\ region.  There is a
separate compact \hii\ region along the eastern border of the diffuse
region, G052.940$-$0.588.  \citet{lockman89} found RRL emission for
this nebula at a velocity of 43.5\,\kms.  The velocity discrepancy
between this compact \hii\ region and the $\sim\,53\,\kms$ off-source
emission is puzzling since both sources are spatially coincident. 
Perhaps the two nebulae are not physically associated at all. 


\subsection{The G52 Complex\label{sec:G52}}
The three \hii\ regions just discussed are part of an even larger
complex that is comprised of at least 7 compact nebulae.  We call this
complex ``G52'' and show this complicated field in
Figure~\ref{fig:G52}.  The green numbers and vectors in the upper left
panel mark the RRL velocities (in \kms) of the two Arecibo HRDS nebulae,
G052.201+0.752 and G052.259+0.700, whereas GBT HRDS nebulae are marked
in cyan.  \citet{watson03} measures a $-$3.0\,\kms RRL velocity for
the yellow-marked nebula, G052.230+0.740, using Arecibo with a $1\arcmin$
beam.  \citet{lockman89} finds a 2.8\,\kms\ velocity for this same
source with the $3\,\arcmin$ NRAO 140 Foot telescope beam.

We adopt here a kinematic distance of 10\,\kpc\ for the G52 complex.
This is based on our \hiea\ measurements for the two Arecibo nebulae,
G052.201+0.752 and G052.259+0.700, that support the $\sim10$\,kpc far
distance.  Furthermore, GBT HRDS G052.160+0.706 is nearby and has a
velocity of 7.9\,\kms\ (Paper~II).  Paper~III found a far kinematic
distance of 10\,\kpc\ for this nebula.  Finally, using the
\citet{lockman89} velocity for G052.230+0.740, \citet{anderson09a}
found strong evidence for the far kinematic distance (which is further
supported by the negative velocity measured by \citealt{watson03}).

The spatial distribution of the compact nebulae in the G52 field
define the border of an even larger structure that we call ``G52L,''
which is outlined by the dashed ellipse in the upper left panel of
Fig.~\ref{fig:G52}.  While there are many observations of the smaller
compact \hii\ regions in the field, due to its faint radio flux G52L
has not yet been been observed in RRL emission.  Based on its MIR and
radio continuum morphology together with the ongoing star formation on
its borders, G52L is almost certainly a large \hii\ region.  

G52L seems to be an evolved \hii\ region that has displaced a significant 
amount of material through its expansion, leaving a low density, low emission
measure internal cavity. 
An \hi\ shell spatially coincident with the photo-dissociation region
(PDR) seen at 8.0\,\micron\ is clearly visible (see
Figure~\ref{fig:G52}, bottom left panel).  The velocity of this shell,
1.8\,\kms, suggests that G52L is associated with the smaller compact
\hii\ regions.  To our knowledge this shell has not been identified
previously as an \hii\ region.  In the bottom right panel of
Figure~\ref{fig:G52}, we show \cor\ emission at 4.2 \kms\ from the
Galactic Ring Survey \citep[GRS; ][]{jackson06}.  There is \cor\ gas
along the north and south PDRs of G52L.  This gas may be tracing
future sites of star formation in the region.  These \hi\ and
\cor\ shells indicate that a significant amount of material has been
displaced by the expansion of G52L.

G52L may be the largest single \hii\ region in the Galaxy.  The PDR
region traced at 8.0\,\micron\ in Figure~\ref{fig:G52} is
$\sim30\,\arcmin$ in diameter.  Using the distance to the compact
sources, $\sim10$\,kpc, this corresponds to a physical diameter of
90\,pc (the \hi\ and \cor\ shells give the same result).  This is
about 40\% the MIR size we measure for the Large Magellanic Cloud's
30~Doradus nebula, which is the largest \hii\ region in the Local
Universe \citep{freedman10}.

 
We estimate the size of various other large, well-known \hii\ regions
using GLIMPSE 8.0\,\micron\ and WISE 12\,\micron\ data\footnote{Both
  data sets trace \hii\ region PDRs so they are equivalent for our
  purposes \citep{anderson12a}.}.  We find that G52L is three times
larger than W43 ($15\,\arcmin$ diameter and 5.7\,kpc distance from
\citealt{anderson09a}), over twice as large as the G305 complex
($30\,\arcmin$ diameter and 4\,kpc distance from \citealt{russeil98};
see recent paper by \citealt{hindson12}), about twice the size of W5-E
($50\,\arcmin$ diameter and assuming the 2.0\,kpc distance for W3 from
\citealt{xu06} also applies to W5; see recent paper L.~Deharveng et al.,
2012, submitted), about twice the size of W5-W ($1\degree$ diameter
and the same 2.0\,kpc distance), about twice as large as CTB~102
($35\,\arcmin$ diameter, not including any associated radio continuum
``filaments,'' and 4.3\,kpc distance from \citealt{arvidsson09}) and
three times larger than W51 ($20\,\arcmin$ diameter of main
\hii\ region W51B from \citealt{westerhaut58} and 5.6\,kpc distance from
\citealt{anderson09a}).

A region as large and energetic as G52L can shape the evolution of
subsequent generations of star formation in its vicinity.  Paper~II 
and \citet{deharveng10} show that nearly all cataloged IR bubbles
enclose \hii\ regions.  G52L is spatially coincident with the infrared
dust bubble N109 in the \citet{churchwell06} compilation while the
smaller loop to the southwest has been separately cataloged as N108.
There is no evidence, however, that N109 and N108 are not part of the
same large source.  Interestingly, most of the small \hii\ regions are
also spatially coincident with infrared dust bubbles: G052.160+0.706
with N110, G052.201+0.752 with N111, G052.230+0.740 with N113, and
G052.259+0.700 with N114\footnote{Notably missing from this list is N112 
which is yet to be observed in RRL emission. Nonetheless, based on its MIR and radio
morphology N112 is probably an \hii\ region.}.  This high density of bubble
\hii\ regions is unusual and may indicate that the G52 local
environment is conducive to their formation.

We hypothesize that there may be many other large, evolved Galactic
\hii\ regions similar to G52L that, due to their low flux densities,
remain undiscovered.  G52L is 10\kpc distant and has a 1.4\ghz flux
density of $\sim11\,\jy$.  This implies an ionizing luminosity of
$\sim\expo{49.9}\,{\rm s^{-1}}$, equivalent to 2 main sequence O4
stars or $\sim$11 main sequence 07 stars \citep{sternberg03}.  Because
it is relatively faint G52L is not in the \citet{murray10} catalog of
large massive star forming regions.  Its exclusion hints that other,
yet to be discovered, luminous diffuse \hii\ regions such as G52L may
make a significant contribution to the total ionization of the Galaxy.


\section{Summary}
Using the Arecibo Observatory at X-band (3\cm), we detect RRL emission
from 37 previously unknown \hii\ regions in the Galactic zone
$66\degree \ge \gl \ge 31\degree$ and $\absb \le 1\degree$.  As in the
GBT HRDS, the Arecibo HRDS candidate \hii\ region targets are selected
based on spatially coincident 24\,\micron\ and 20\,cm emission that
has a similar morphology and angular extent.  Compared to the GBT
HRDS, the Arecibo HRDS targets are fainter, smaller in angular size,
or are in more crowded fields.  The Arecibo HRDS nebulae are among the
faintest ever detected in RRL emission.  Only 58\% of the Arecibo HRDS
candidate \hii regions show RRL emission whereas the detection rate
for GBT HRDS targets was 95\%.
We do not really know why the discovery rate for Arecibo \hii region
candidates is so low compared with that for the GBT HRDS.  It may just
be the case that the relatively low detection rate of Arecibo \hii
region candidates is a direct result of the low mean flux of the
Arecibo HRDS nebulae.  Despite having an average spectral sensitivity
a factor of two better, the mean signal-to-noise ratio (SNR) for
Arecibo \hii region detections is 5.0 compared to the 17.7 SNR for GBT
HRDS nebulae.  Perhaps the GBT census had already detected the bulk of
the \hii regions in this part of the Milky Way that are detectable by
Arecibo with reasonable integration times.  We also may be finding the
first indication of a real drop off in the \hii region luminosity
distribution.

%

The Arecibo HRDS enhances the census of regions of massive star
formation for that part of the first quadrant Galactic disk that is
accessible to the Arecibo telescope.  This census reveals five
\hii\ regions near $\gl\,\sim\,50\degree$ that have velocities $\gsim
20 - 25 \kms$\ beyond the tangent point velocity in this direction.  
This is evidence for strong streaming motions at this location.  Over
the Arecibo longitude range, the Galactic radial distribution of
\hii\ regions shows peaks at $\sim6$\,kpc and $\sim8$\,kpc.  The
8\,kpc peak for the Arecibo HRDS is more prominent than that for the
distribution of \hii\ regions known previously.

We derive kinematic distances for 23 Arecibo HRDS nebulae.  Four of
these have negative LSR velocities, placing them unambiguously in
the outer Galaxy.  Of the remaining sources, 12 are at the tangent
point distance and the others are at the far kinematic distance.  

We find, apparently for the first time, a large \hii\ region, G52L,
that is a member of a complex of \hii\ regions 10\,kpc distant.  G52L
may be the largest single \hii\ region in the Galaxy.  It is
physically larger than all the well-known \hii\ regions we discuss
here and is about 40\% the size of 30~Doradus in the Large Magellanic
Cloud.  G52L is associated with a $30\,\arcmin$ diameter \hi\ shell that
is spatially coincident with its PDR, and also with \cor\ emission on
its northern and southern boundaries.  Together with most of the
\hii\ regions in the complex, G52L has a bubble morphology when
observed at mid-infrared wavelengths.  This concentration of bubbles
is unusual and may indicate something about the local environment.


\appendix
\section{The HRDS Web Site}
\begin{figure*}[!ht]
  \centering
  \includegraphics[width=6.5 in]{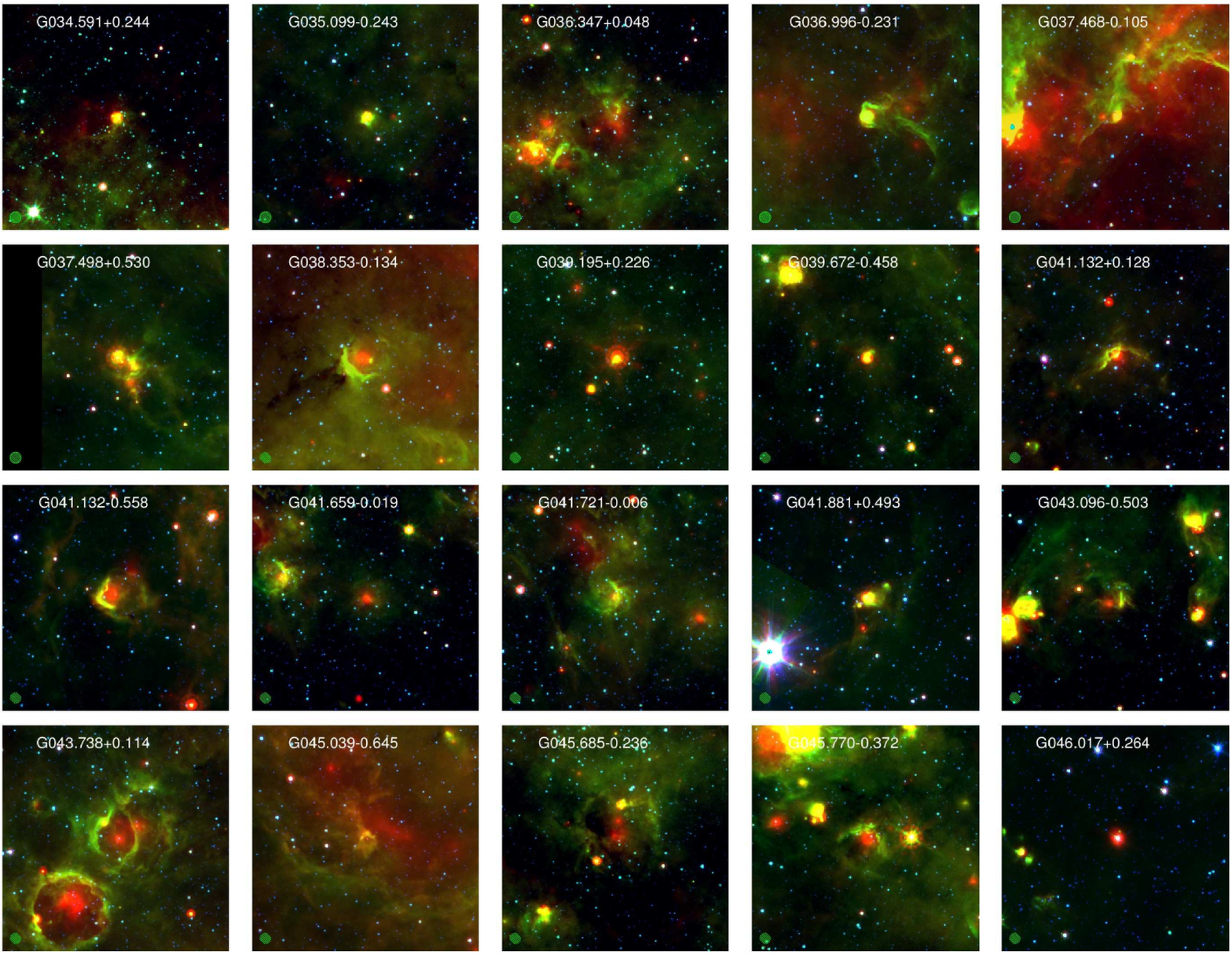}
  \caption{ {\em Spitzer} infrared images of Arecibo HRDS nebulae.
    Image data are from MIPSGAL 24\,\micron\ (red), GLIMPSE
    8.0\,\micron\ (green) and GLIMPSE 3.6\,\micron\ (blue).  Each
    image is $12\,\arcmin$ square; the $30\,\arcsec$ Arecibo beam is
    shown in the lower left corner.}
 \label{fig:three_color}
\end{figure*}

All the Arecibo HRDS data are now incorporated in our HRDS Web site,
\url{http://go.nrao.edu/hrds}, where one can view the three-color
images (Fig.~\ref{fig:three_color}) and the \hnaa\ recombination line
spectra (Fig.~\ref{fig:spectra}).  One can also download the contents
of Tables~\ref{tab:sources}, \ref{tab:line}, and \ref{tab:dist}.  As
we continue to extend the HRDS all future data will also be available
on this site.

\begin{acknowledgments}
The Arecibo Observatory is operated by SRI International under a
cooperative agreement with the National Science Foundation
(AST-1100968), and in alliance with Ana G. M\'endez-Universidad
Metropolitana, and the Universities Space Research Association.  This
research has made use of NASA's Astrophysics Data System Bibliographic
Services and the SIMBAD database operated at CDS, Strasbourg, France.

\it Facility: Arecibo Observatory
\end{acknowledgments}

\bibliographystyle{apj}
\bibliography{/home/bania/papers/bozo/paper/ref}
\clearpage

\begin{figure*}[!ht]
  \centering
  \includegraphics[width=6.5 in]{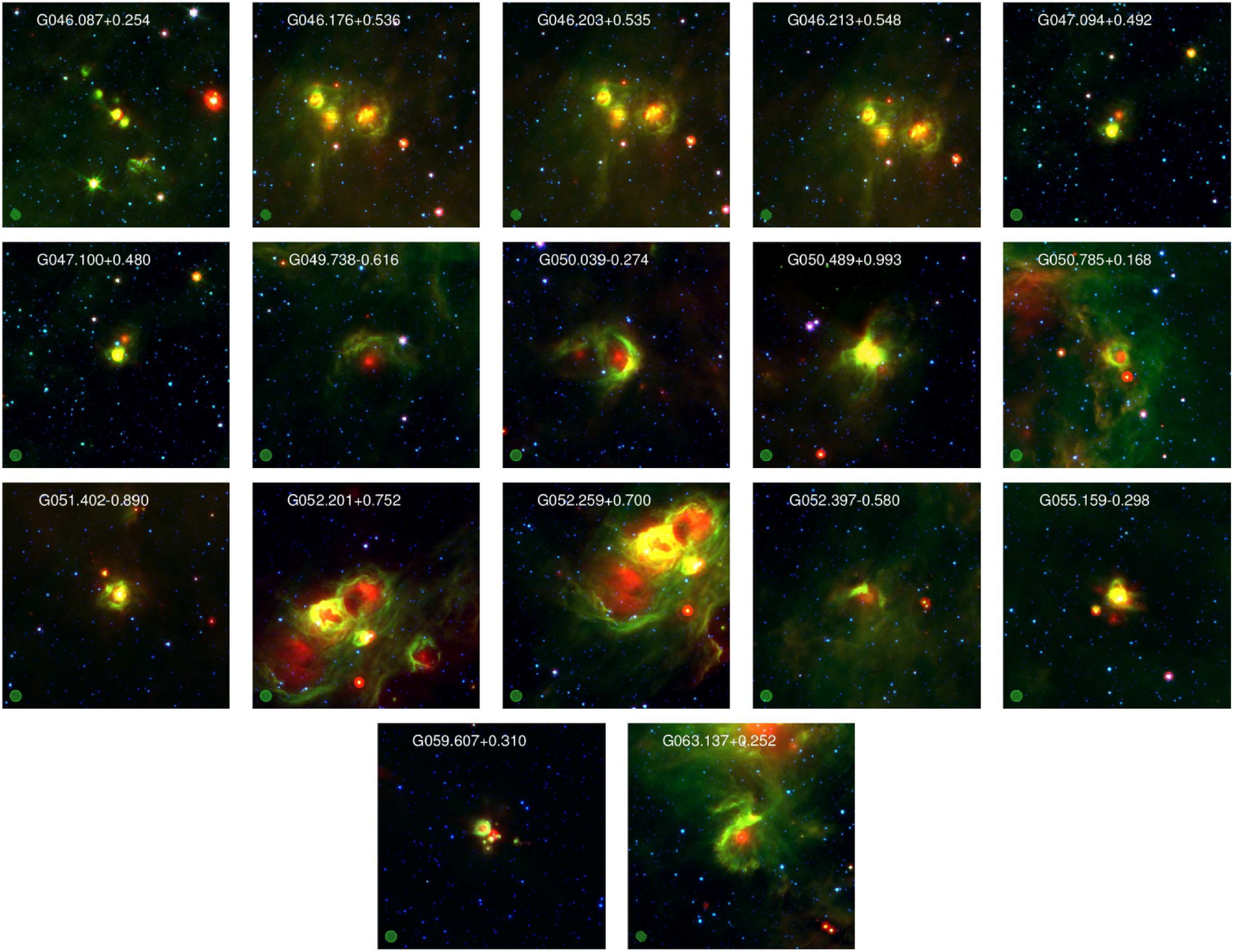}
\end{figure*}
\clearpage

\begin{figure}
  \centering
  \includegraphics[width=6 in]{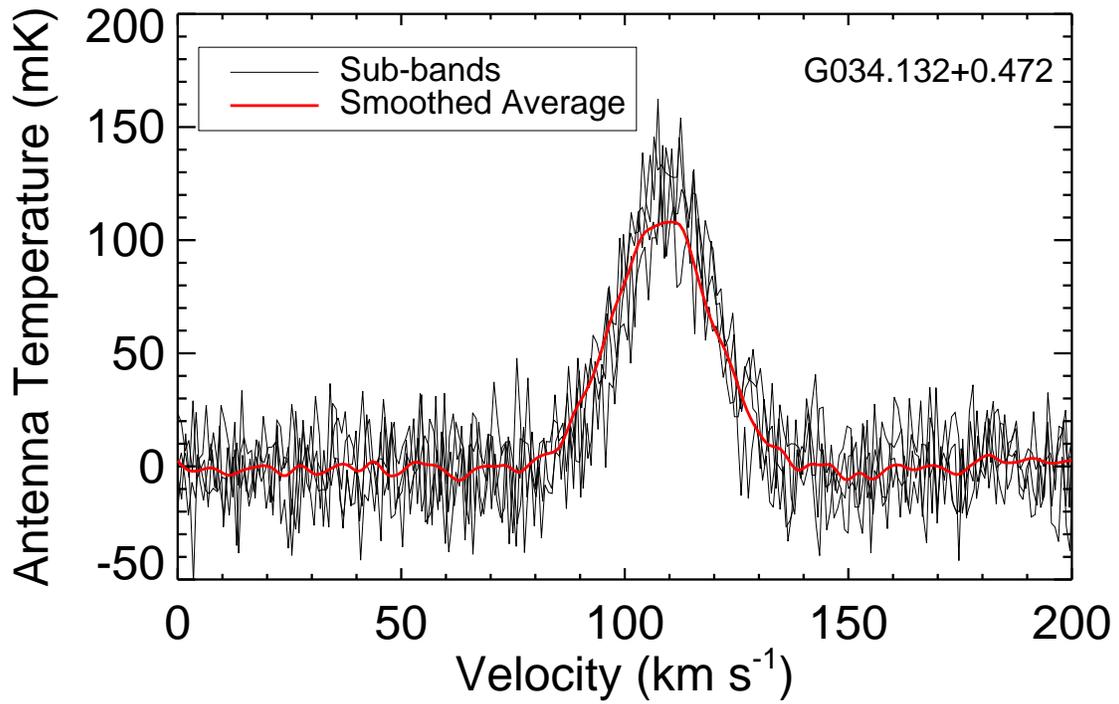}
  \caption{G034.132+0.472 \hii\ region spectra.  Shown in black are
    spectra for the four different tunings (sub-bands).  Each is the
    average of both circular polarizations.  The red line \hnaa\  
    spectrum is the average of these data smoothed to 
    4.2\,\kms\ resolution.}
  \label{fig:four_ifs}
\end{figure}
\clearpage

\begin{figure*}[!ht]
  \centering
  \includegraphics[width=6.5 in]{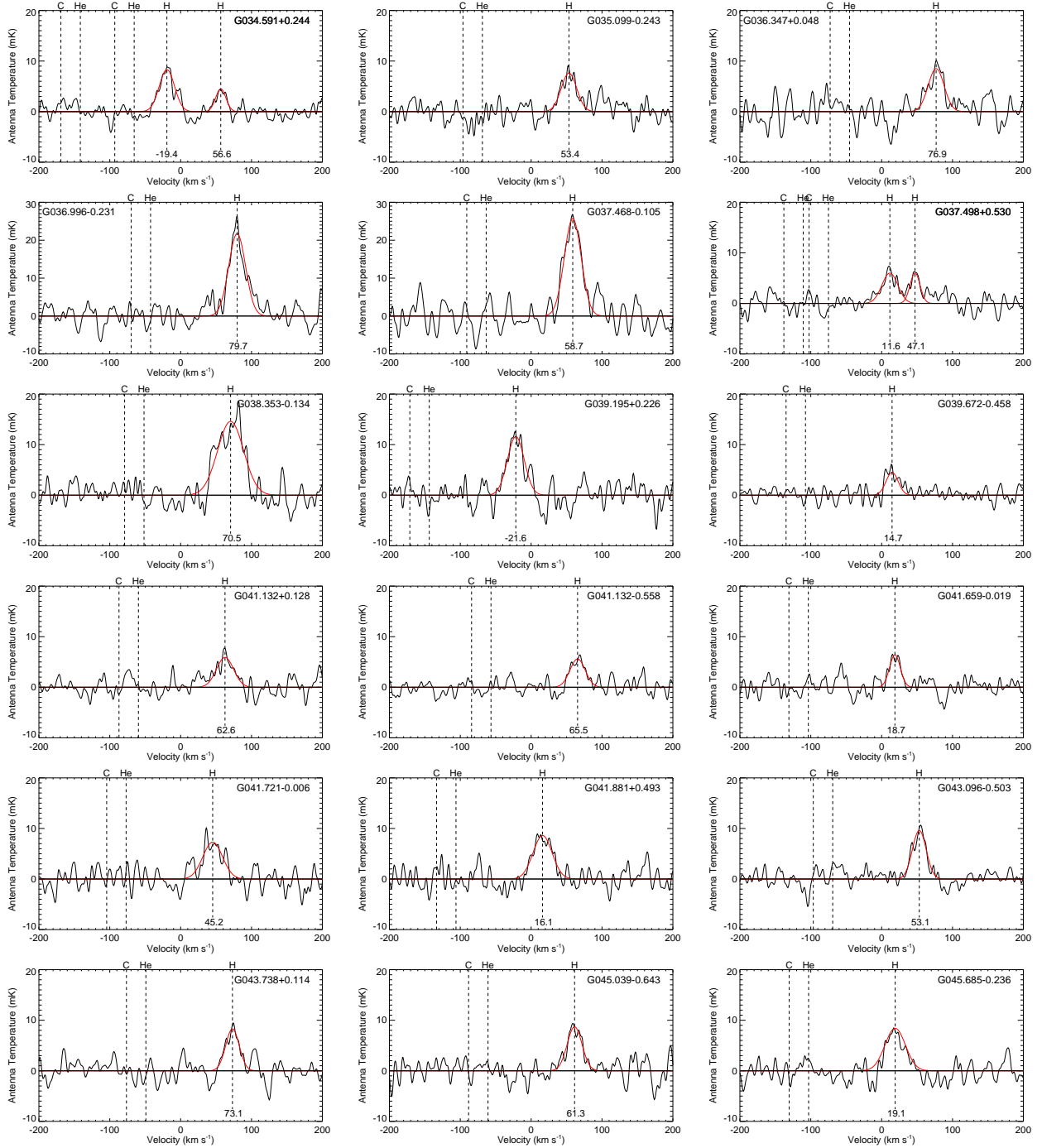}
  \caption{Arecibo HRDS composite \hnaa\ RRL discovery spectra.  
    Shown is the average of the \hal{89} through \hal{92} RRLs,
    smoothed to 4.2\,\kms\ resolution.  A Gaussian fit to each
    hydrogen RRL component is superimposed in red.  A vertical dashed line
    flags the nebula's hydrogen RRL LSR velocity, which is listed at
    the bottom of the flag.  The expected locations of the helium and
    carbon RRLs are also flagged.}
  \label{fig:spectra}
\end{figure*}

\begin{figure*}[!ht]
  \centering
  \includegraphics[width=6.5 in]{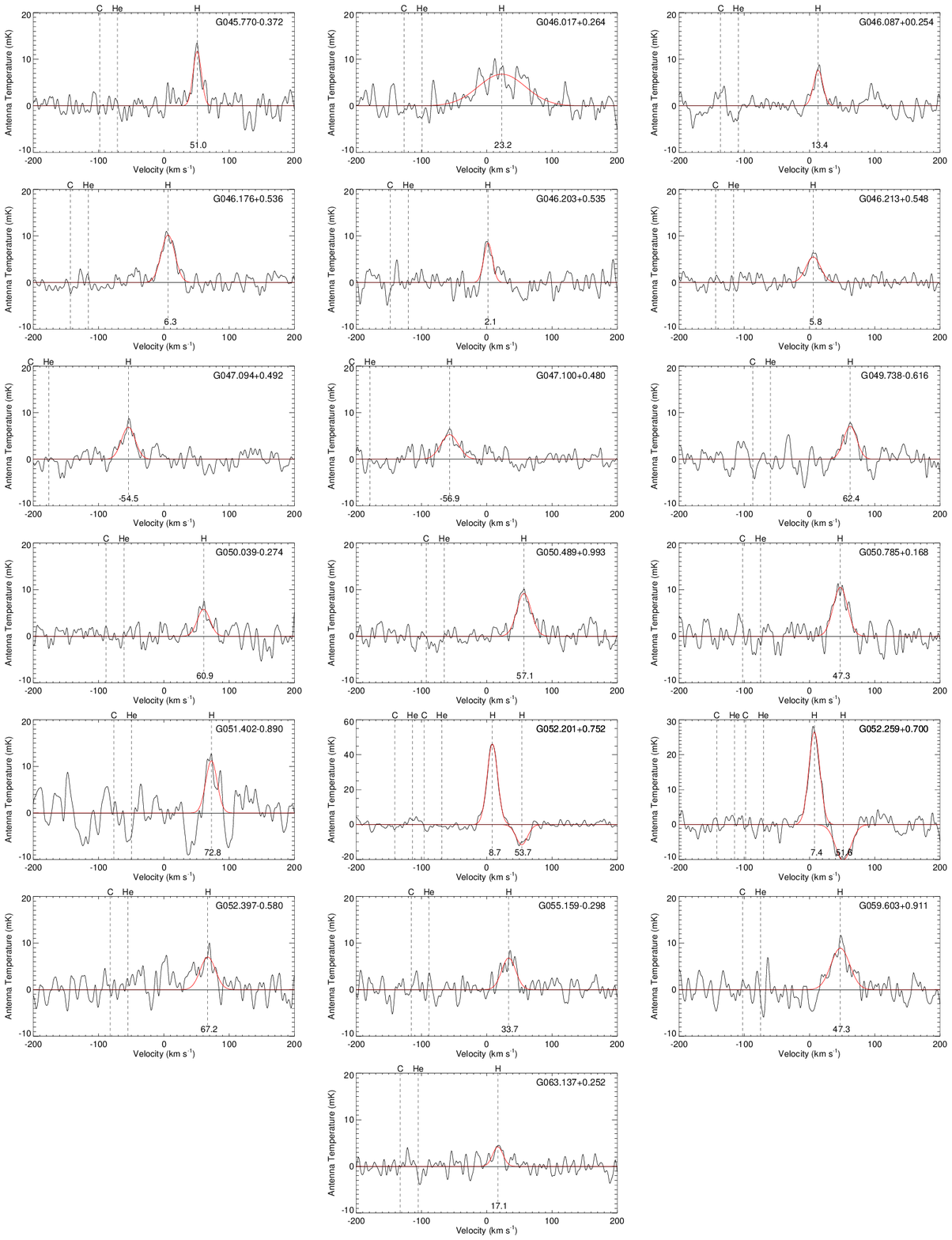}
\end{figure*}

\clearpage

\begin{figure}
  \centering
  \includegraphics[width=6 in]{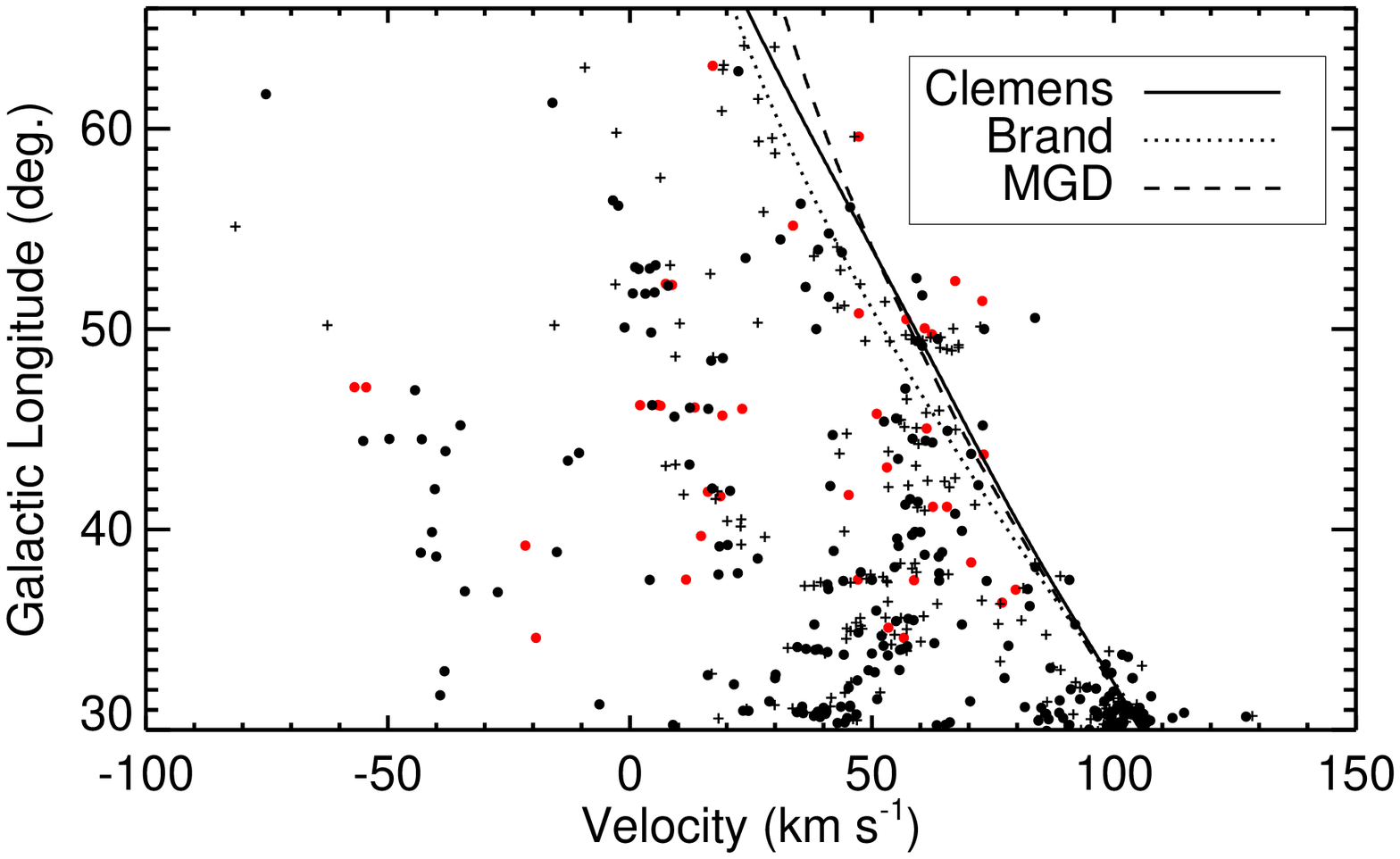}
  \caption{\hii\ region longitude-velocity diagram for the first
    quadrant Galactic plane visible to the Arecibo telescope.  Shown
    is the current census of all known \hii\ regions in this zone
    from: the Arecibo HRDS (red circles), the GBT HRDS (black
    circles), and nebulae known prior to the GBT HRDS (black crosses;
    see Paper~II).  Tangent point velocity loci for the
    \citet{clemens85}, \citet{brand86}, and \citet{mcclure07} rotation
    curves are drawn as solid, dotted, and dashed lines, respectively.
    In the longitude range $\gl\,\sim\,50\degree\, - \,52\degree$
    there are 18 nebulae with velocities that are beyond the terminal
    velocities predicted by all three of the rotation curve models.  }
  \label{fig:lv}
\end{figure}
\clearpage

\begin{figure}
  \centering
  \includegraphics[width=6 in]{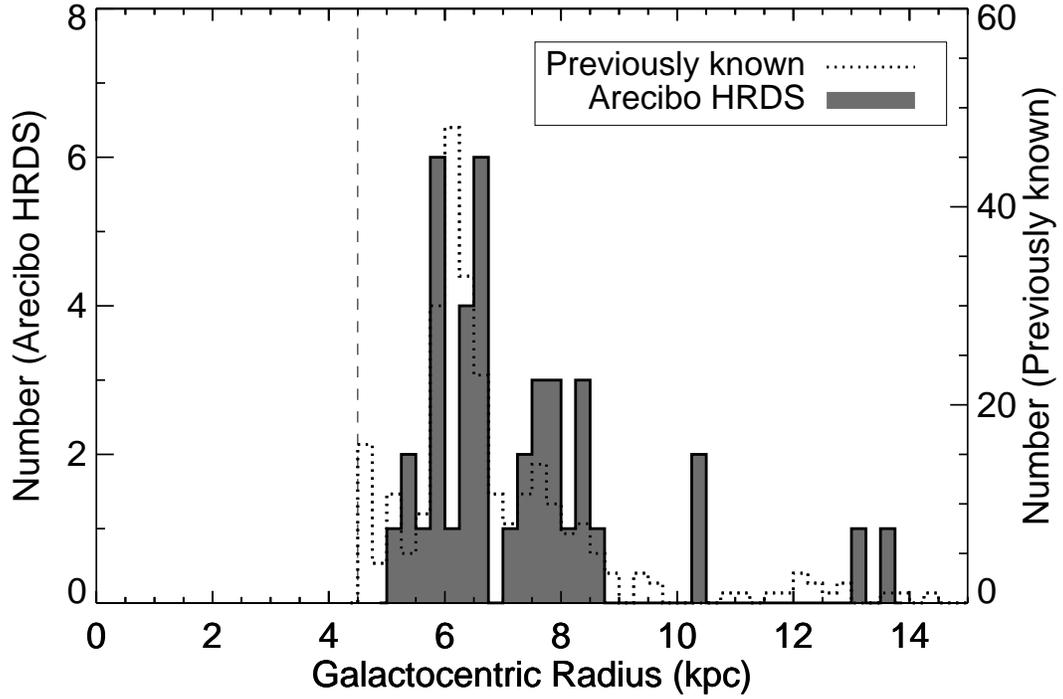}
  \caption{Galactic radial distribution of \hii\ regions.  Arecibo
    HRDS nebulae are the gray filled histogram and the previously
    known sample of \hii\ regions, including the GBT HRDS, is the
    dotted line histogram.  Only nebulae located in the longitude
    range of the Arecibo HRDS are included here.  The vertical dashed
    line shows the minimum Galactocentric radius sampled by the
    present study.}
  \label{fig:rgal}
\end{figure}
\clearpage
\begin{figure}
  \centering
  \includegraphics[width=6.2 in]{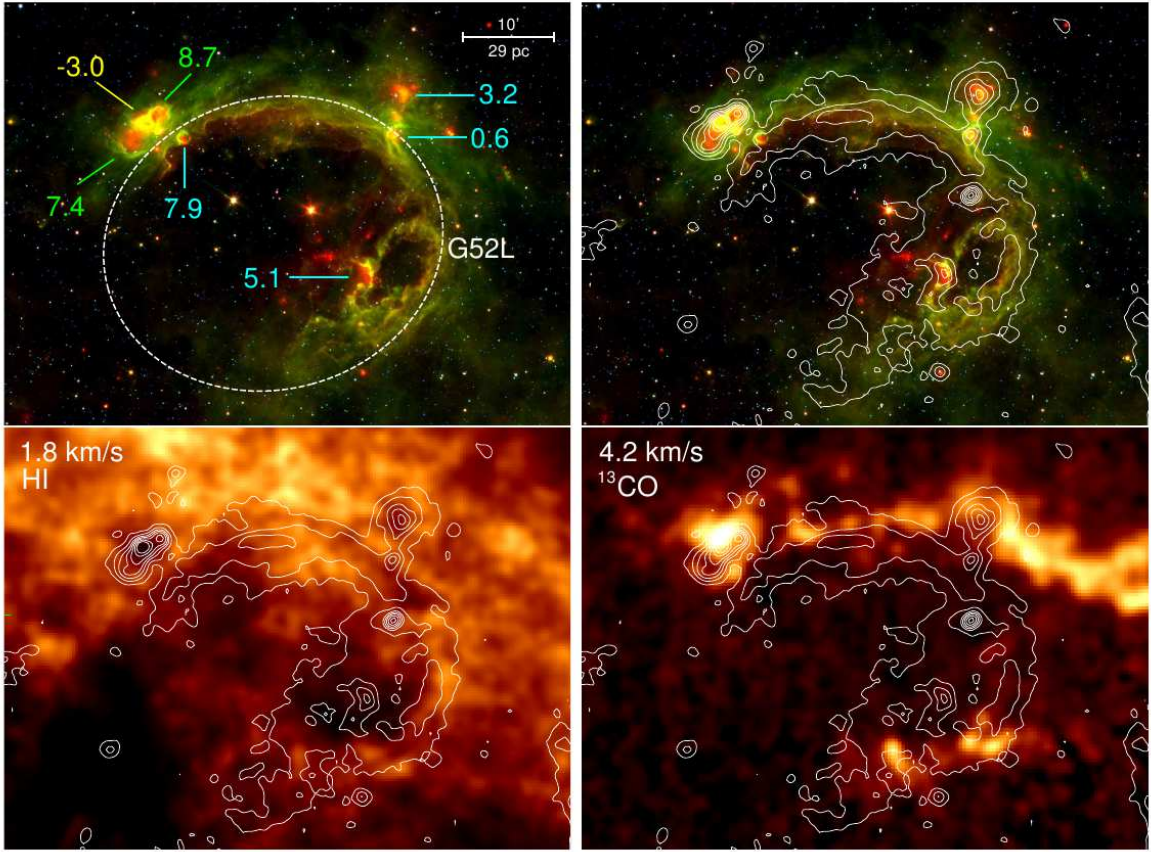}
  \caption{Multi-wavelength images of the G52 complex.  Each is
    centered at \lb\ = ($51\,\fdg97, 0\,\fdg57$) and shows a region
    spanning $1\degree\times~0\,\fdg75$ in Galactic co-ordinates.  The
    scale bar in the top left panel shows $10\,\arcmin$, which is
    29\,pc at a distance of 10\,kpc; the entire field shown is $175
    \times 130$\,pc. G52L is the large half-circle structure open to
    the south-east, marked with a white dashed ellipse in the top left
    panel.  Contours are of VGPS 21\,cm continuum emission at levels
    of 9, 12, 16, 20, 25, 32, and 40\,K.  Background images for the
    top two panels are {\it Spitzer} data: MIPSGAL 24\,\micron\ (red),
    GLIMPSE 8.0\,\micron\ (green) and GLIMPSE 3.6\,\micron\ (blue).
    In the top left panel, the RRL velocities (in \kms) and vectors point
    to different nebulae in this complex.  Colors identify the source
    of these data: green for this work, cyan for the GBT HRDS, and
    yellow for \citet{watson03}.  The bottom left panel shows a single
    \hi\ channel from the VGPS at a velocity of 1.8\,\kms.  The bottom
    right panel shows a single \cor\ channel from the GRS at a
    velocity of 4.2\,\kms. Both bottom panel images are spatially
    smoothed with a $1\arcmin$ FWHM Gaussian kernel.}
  \label{fig:G52}
\end{figure}
\clearpage

\begin{deluxetable}{lcccccc}
\tabletypesize{\scriptsize}
\tablecaption{Arecibo HRDS Source Catalog}
\tablewidth{0pt}
\tablehead{
\colhead{Name} &
\colhead{\gl} &
\colhead{\gb} &
\colhead{RA(J2000)} &
\colhead{Dec.(J2000)} &
\colhead{CS\tablenotemark{a}} &
\colhead{Morphology\tablenotemark{b}}
\\
\colhead{} &
\colhead{deg} &
\colhead{deg} &
\colhead{hh:mm:ss.s} &
\colhead{dd:mm:ss} &
\colhead{\kms} &
\colhead{}
}
\startdata
G034.591+0.244 & 34.591  &  \phs 0.244  &  18:53:35.9  &  +01:35:17  &  \nodata  &  C  \\
G035.099$-$0.243 & 35.099  &  $-0.243$  &  18:56:15.5  &  +01:49:04  &  \nodata  &  C  \\
G036.347+0.048 & 36.347  &  \phs 0.048  &  18:57:30.0  &  +03:03:39  &  \nodata  &  I  \\
G036.996$-$0.231 & 36.996  &  $-0.231$  &  18:59:41.1  &  +03:30:39  &  \nodata  &  C  \\
G037.468$-$0.105 & 37.468  &  $-0.105$  &  19:00:05.9  &  +03:59:19  &  \nodata  &  C  \\
G037.498+0.530 & 37.498  &  \phs 0.530  &  18:57:53.4  &  +04:18:17  &  \nodata  &  C  \\
G038.353$-$0.134 & 38.353  &  $-0.134$  &  19:01:49.7  &  +04:45:43  &  ND  &  B (N72)  \\
G039.195+0.226 & 39.195  &  \phs 0.226  &  19:02:05.6  &  +05:40:31  &  \phn2.6  &  C  \\
G039.672$-$0.458 & 39.672  &  $-0.458$  &  19:05:24.7  &  +05:47:09  &  ND  &  C  \\
G041.132+0.128 & 41.132  &  \phs 0.128  &  19:06:01.0  &  +07:21:05  &  \nodata  &  PB  \\
G041.132$-$0.558 & 41.132  &  $-0.558$  &  19:08:28.4  &  +07:02:11  &  \nodata  &  B  \\
G041.659$-$0.019 & 41.659  &  $-0.019$  &  19:07:31.5  &  +07:45:07  &  \nodata  &  IB  \\
G041.721$-$0.006 & 41.721  &  $-0.006$  &  19:07:35.2  &  +07:48:45  &  \nodata  &  I  \\
G041.881+0.493 & 41.881  &  \phs 0.493  &  19:06:05.7  &  +08:11:04  &  \nodata  &  C  \\
G043.096$-$0.503 & 43.096  &  $-0.503$  &  19:11:56.3  &  +08:48:11  &  \nodata  &  I  \\
G043.738+0.114 & 43.738  &  \phs 0.114  &  19:10:55.5  &  +09:39:27  &  \nodata  &  B (N89)  \\
G045.039$-$0.643 & 45.039  &  $-0.643$  &  19:16:06.0  &  +10:27:35  &  \nodata  &  I  \\
G045.685$-$0.236 & 45.685  &  $-0.236$  &  19:15:51.7  &  +11:13:12  &  \nodata  &  IB  \\
G045.770$-$0.372 & 45.770  &  $-0.372$  &  19:16:30.7  &  +11:13:55  &  \nodata  &  B  \\
G046.017+0.264 & 46.017  &  \phs 0.264  &  19:14:41.0  &  +11:44:50  &  \nodata  &  PS  \\
G046.087+0.254 & 46.087  &  \phs 0.254  &  19:14:51.2  &  +11:48:15  &  \nodata  &  C  \\
G046.176+0.536 & 46.176  &  \phs 0.536  &  19:13:60.0  &  +12:00:50  &  \nodata  &  B  \\
G046.203+0.535 & 46.203  &  \phs 0.535  &  19:14:03.5  &  +12:02:13  &  \nodata  &  IB  \\
G046.213+0.548 & 46.213  &  \phs 0.548  &  19:14:01.7  &  +12:03:10  &  \nodata  &  IB  \\
G047.094+0.492 & 47.094  &  \phs 0.492  &  19:15:54.6  &  +12:48:24  &  ND  &  B  \\
G047.100+0.480 & 47.100  &  \phs 0.480  &  19:15:57.8  &  +12:48:21  &  ND  &  C  \\
G049.738$-$0.616 & 49.738  &  $-0.616$  &  19:25:03.0  &  +14:37:11  &  \nodata  &  B  \\
G050.039$-$0.274 & 50.039  &  $-0.274$  &  19:24:23.5  &  +15:02:49  &  \nodata  &  B (N104)  \\
G050.489+0.993 & 50.489  &  \phs 0.993  &  19:20:38.2  &  +16:02:29  &  \nodata  &  BB  \\
G050.785+0.168 & 50.785  &  \phs 0.168  &  19:24:14.6  &  +15:54:48  &  \nodata  &  B (N106)  \\
G051.402$-$0.890 & 51.402  &  $-0.890$  &  19:29:20.1  &  +15:57:07  &  \nodata  &  B  \\
G052.201+0.752 & 52.201  &  \phs 0.752  &  19:24:53.7  &  +17:26:15  &  \nodata  &  B  \\
G052.259+0.700 & 52.259  &  \phs 0.700  &  19:25:12.1  &  +17:27:48  &  \nodata  &  B  \\
G052.397$-$0.580 & 52.397  &  $-0.580$  &  19:30:11.5  &  +16:58:27  &  \nodata  &  IB  \\
G055.159$-$0.298 & 55.159  &  $-0.298$  &  19:34:45.7  &  +19:31:46  &  41.1  &  C  \\
G059.603+0.911 & 59.603  &  \phs 0.911  &  19:39:34.8  &  +23:59:47  &  36.9  &  B  \\
G063.137+0.252 & 63.137  &  \phs 0.252  &  19:49:56.0  &  +26:43:39  &  \nodata  &  IB  \\

\enddata
\tablenotetext{a}{From \citet{bronfman96};  ``ND'' means observed CS positions that lack emission.}\\
\tablenotetext{b}{Morphological source structure classification based on {\it Spitzer} GLIMPSE 8\micron\ images:\\  
B -- Bubble: 8\,\micron\ emission surrounding 24\,\micron\ and radio continuum emission\\
BB -- Bipolar Bubble:  two bubbles connected by a region of strong IR and radio continuum emission\\
PB -- Partial Bubble:  similar to ``B'' but not complete\\
IB -- Irregular Bubble:  similar to ``B'' but with less well-defined structure\\ 
C -- Compact:  resolved 8\,\micron\ emission with no hole in the center\\
PS -- Point Source:  unresolved 8\,\micron\ emission\\
I -- Irregular:  complex morphology not easily classified\\
We list in parenthesis the bubble identification name for sources in
the \citet{churchwell06} IR bubble catalog.}
\label{tab:sources}
\end{deluxetable}
\clearpage

\begin{deluxetable}{lccrcccrcc}
\tabletypesize{\scriptsize}
\tablecaption{Arecibo HRDS Hydrogen Recombination Line Parameters}
\tablewidth{0pt}
\tablehead{
\colhead{Name} &
\colhead{\gl} &
\colhead{\gb} &
\colhead{$T_L$} &
\colhead{$\sigma T_L$} &
\colhead{$\Delta V$} &
\colhead{$\sigma \Delta V$} &
\colhead{$V_{LSR}$} &
\colhead{$\sigma V_{LSR}$} &
\colhead{rms}
\\
\colhead{} &
\colhead{(deg)} &
\colhead{(deg)} &
\colhead{(mK)} &
\colhead{(mK)} &
\colhead{(\kms)} &
\colhead{(\kms)} &
\colhead{(\kms)} &
\colhead{(\kms)} &
\colhead{(mK)}
}
\startdata

G034.591+0.244a & 34.591  &  \phs 0.244  &  \phn 4.5  &  0.4  &  17.0  &  1.6  &  \phs 56.6  &  0.7 & 1.2 \\
G034.591+0.244b & 34.591  &  \phs 0.244  &  \phn 8.3  &  0.3  &  25.4  &  1.1  &  $-19.4$  &  0.5 & 1.2 \\
G035.099$-$0.243 & 35.099  &  $-0.243$  &  \phn 7.6  &  0.3  &  25.8  &  1.2  &  \phs 53.4  &  0.5 & 1.8 \\
G036.347+0.048 & 36.347  &  \phs 0.048  &  \phn 8.5  &  0.4  &  26.5  &  1.4  &  \phs 76.9  &  0.6 & 2.1 \\
G036.996$-$0.231 & 36.996  &  $-0.231$  &  21.7  &  0.6  &  28.0  &  0.9  &  \phs 79.7  &  0.4 & 2.1 \\
G037.468$-$0.105 & 37.468  &  $-0.105$  &  25.7  &  0.5  &  28.0  &  0.6  &  \phs 58.7  &  0.3 & 3.3 \\
G037.498+0.530a & 37.498  &  \phs 0.530  &  \phn 6.0  &  0.3  &  16.6  &  0.9  &  \phs 47.1  &  0.4 & 1.5 \\
G037.498+0.530b & 37.498  &  \phs 0.530  &  \phn 6.0  &  0.2  &  26.4  &  1.2  &  \phs 11.6  &  0.5 & 1.5 \\
G038.353$-$0.134 & 38.353  &  $-0.134$  &  14.6  &  0.6  &  43.2  &  2.0  &  \phs 70.5  &  0.8 & 1.9 \\
G039.195+0.226 & 39.195  &  \phs 0.226  &  11.6  &  0.5  &  27.9  &  1.3  &  $-21.6$  &  0.6 & 2.1 \\
G039.672$-$0.458 & 39.672  &  $-0.458$  &  \phn 4.5  &  0.4  &  20.2  &  2.4  &  \phs 14.7  &  1.0 & 1.4 \\
G041.132+0.128 & 41.132  &  \phs 0.128  &  \phn 5.9  &  0.3  &  28.4  &  1.7  &  \phs 62.6  &  0.7 & 1.4 \\
G041.132$-$0.558 & 41.132  &  $-0.558$  &  \phn 5.6  &  0.2  &  25.3  &  0.9  &  \phs 65.5  &  0.4 & 1.3 \\
G041.659$-$0.019 & 41.659  &  $-0.019$  &  \phn 6.4  &  0.3  &  17.9  &  0.9  &  \phs 18.7  &  0.4 & 1.4 \\
G041.721$-$0.006 & 41.721  &  $-0.006$  &  \phn 7.3  &  0.4  &  32.6  &  2.3  &  \phs 45.2  &  0.9 & 1.8 \\
G041.881+0.493 & 41.881  &  \phs 0.493  &  \phn 8.6  &  0.2  &  31.6  &  1.1  &  \phs 16.1  &  0.4 & 1.7 \\
G043.096$-$0.503 & 43.096  &  $-0.503$  &  \phn 9.6  &  0.3  &  23.0  &  1.0  &  \phs 53.1  &  0.4 & 1.5 \\
G043.738+0.114 & 43.738  &  \phs 0.114  &  \phn 8.3  &  0.3  &  22.4  &  0.9  &  \phs 73.1  &  0.4 & 1.7 \\
G045.039$-$0.643 & 45.039  &  $-0.643$  &  \phn 8.7  &  0.3  &  24.6  &  0.9  &  \phs 61.3  &  0.4 & 1.8 \\
G045.685$-$0.236 & 45.685  &  $-0.236$  &  \phn 8.4  &  0.3  &  34.0  &  1.3  &  \phs 19.1  &  0.5 & 1.6 \\
G045.770$-$0.372 & 45.770  &  $-0.372$  &  11.7  &  0.6  &  15.1  &  0.9  &  \phs 51.0  &  0.4 & 1.9 \\
G046.017+0.264 & 46.017  &  \phs 0.264  &  \phn 6.8  &  0.3  &  85.2  &  4.4  &  \phs 23.2  &  1.7 & 2.2 \\
G046.087+0.254 & 46.087  &  \phs 0.254  &  \phn 7.6  &  0.3  &  17.0  &  0.9  &  \phs 13.4  &  0.4 & 1.8 \\
G046.176+0.536 & 46.176  &  \phs 0.536  &  10.2  &  0.2  &  24.8  &  0.6  &  \phn \phs 6.3  &  0.2 & 1.2 \\
G046.203+0.535 & 46.203  &  \phs 0.535  &  \phn 8.5  &  0.6  &  15.7  &  1.2  &  \phn \phs 2.1  &  0.5 & 1.9 \\
G046.213+0.548 & 46.213  &  \phs 0.548  &  \phn 5.5  &  0.2  &  26.3  &  1.3  &  \phn \phs 5.8  &  0.6 & 1.4 \\
G047.094+0.492 & 47.094  &  \phs 0.492  &  \phn 6.9  &  0.3  &  24.8  &  1.3  &  $-54.5$  &  0.5 & 1.4 \\
G047.100+0.480 & 47.100  &  \phs 0.480  &  \phn 5.3  &  0.2  &  31.6  &  1.7  &  $-56.9$  &  0.7 & 1.4 \\
G049.738$-$0.616 & 49.738  &  $-0.616$  &  \phn 7.1  &  0.3  &  23.9  &  1.2  &  \phs 62.4  &  0.5 & 2.1 \\
G050.039$-$0.274 & 50.039  &  $-0.274$  &  \phn 5.8  &  0.4  &  22.7  &  1.7  &  \phs 60.9  &  0.7 & 1.9 \\
G050.489+0.993 & 50.489  &  \phs 0.993  &  \phn 9.2  &  0.3  &  25.9  &  1.1  &  \phs 57.1  &  0.5 & 1.5 \\
G050.785+0.168 & 50.785  &  \phs 0.168  &  10.5  &  0.4  &  25.8  &  1.2  &  \phs 47.3  &  0.5 & 2.1 \\
G051.402$-$0.890 & 51.402  &  $-0.890$  &  11.2  &  0.8  &  20.5  &  0.2  &  \phs 72.8  &  0.7 & 3.5 \\
G052.201+0.752 & 52.201  &  \phs 0.752  &  46.1  &  0.5  &  19.0  &  0.2  &  \phn \phs 8.7  &  0.1 & 1.6 \\
G052.259+0.700 & 52.259  &  \phs 0.700  &  26.5  &  0.4  &  20.0  &  0.4  &  \phn \phs 7.4  &  0.1 & 1.9 \\
G052.397$-$0.580 & 52.397  &  $-0.580$  &  \phn 7.0  &  0.4  &  27.4  &  2.3  &  \phs 67.2  &  0.8 & 1.9 \\
G055.159$-$0.298 & 55.159  &  $-0.298$  &  \phn 6.8  &  0.4  &  25.3  &  1.7  &  \phs 33.7  &  0.7 & 1.7 \\
G059.603+0.911 & 59.603  &  \phs 0.911  &  \phn 8.9  &  0.3  &  34.1  &  1.6  &  \phs 47.3  &  0.6 & 2.1 \\
G063.137+0.252 & 63.137  &  \phs 0.252  &  \phn 4.2  &  0.4  &  18.7  &  1.8  &  \phs 17.1  &  0.8 & 1.3 \\

\enddata
\label{tab:line}
\end{deluxetable}
\clearpage

\begin{deluxetable}{lcccc}
\tabletypesize{\scriptsize}
\tablecaption{Targets not Detected in RRL Emission}
\tablewidth{0pt}
\tablehead{
\colhead{Name} &
\colhead{$l$} &
\colhead{$b$} &
\colhead{rms} &
\colhead{Morphology\tablenotemark{a}} \\
\colhead{} &
\colhead{(deg)} &
\colhead{(deg)} &
\colhead{(mK)} &
\colhead{}
}
\startdata

G031.727+0.699 & 31.727  &  \phs 0.699  &  2.0  &  B  \\
G034.047+0.139 & 34.047  &  \phs 0.139  &  2.3  &  PB  \\
G037.319+0.162 & 37.319  &  \phs 0.162  &  2.2  &  C  \\
G038.906$-$0.437 & 38.906  &  $-0.437$  &  2.4  &  B (N74)  \\
G038.977$-$0.269 & 38.977  &  $-0.269$  &  2.1  &  I  \\
G039.271+0.347 & 39.271  &  \phs 0.347  &  1.9  &  C  \\
G039.506$-$0.280 & 39.506  &  $-0.280$  &  2.2  &  C  \\
G040.017$-$0.119 & 40.017  &  $-0.119$  &  2.0  &  C  \\
G040.422$-$0.039 & 40.422  &  $-0.039$  &  2.6  &  B (N77)  \\
G041.229+0.170 & 41.229  &  \phs 0.170  &  2.2  &  B  \\
G044.244$-$0.129 & 44.244  &  $-0.129$  &  1.9  &  B  \\
G044.353+0.443 & 44.353  &  \phs 0.443  &  2.2  &  I  \\
G044.775$-$0.548 & 44.775  &  $-0.548$  &  3.7  &  B (N93)  \\
G046.060+0.220 & 46.060  &  \phs 0.220  &  2.8  &  I  \\
G047.183+0.317 & 47.183  &  \phs 0.317  &  2.0  &  I  \\
G050.900+1.056 & 50.900  &  \phs 1.056  &  1.8  &  I  \\
G056.252$-$0.160 & 56.252  &  $-0.160$  &  2.7  &  C  \\
G056.255$-$0.099 & 56.255  &  $-0.099$  &  2.8  &  I  \\
G057.469+0.304 & 57.469  &  \phs 0.304  &  1.9  &  IB  \\
G058.121+0.911 & 58.121  &  \phs 0.911  &  1.9  &  C  \\
G058.606+0.620 & 58.606  &  \phs 0.620  &  1.5  &  PB  \\
G059.607+0.310 & 59.607  &  \phs 0.310  &  2.8  &  IB (N126)  \\
G059.885+0.759 & 59.885  &  \phs 0.759  &  2.2  &  B  \\
G060.653$-$0.016 & 60.653  &  $-0.016$  &  1.9  &  PB (N127)  \\
G062.370$-$0.540 & 62.370  &  $-0.540$  &  2.0  &  PB (N130)  \\
G062.723+0.619 & 62.723  &  \phs 0.619  &  1.7  &  IB  \\

\enddata
\tablenotetext{a}{Source structure classification as in 
Table~\ref{tab:sources}. }
\label{tab:nondetections}
\end{deluxetable}
\clearpage

\begin{deluxetable}{llcccccccc}
\tabletypesize{\scriptsize}
\tablecaption{Comparison of Arecibo and GBT HRDS RRL Parameters}
\tablewidth{0pt}
\tablehead{
\colhead{Name} &
\colhead{Telescope\tablenotemark{a}} &
\colhead{$T_L$} &
\colhead{$\sigma T_L$} &
\colhead{$\Delta V$} &
\colhead{$\sigma \Delta V$} &
\colhead{$V_{LSR}$} &
\colhead{$\sigma V_{LSR}$} &
\colhead{rms}
\\
\colhead{} &
\colhead{} &
\colhead{mK} &
\colhead{mK} &
\colhead{\kms} &
\colhead{\kms} &
\colhead{\kms} &
\colhead{\kms} &
\colhead{mK}
}
\startdata
G034.133+0.471 &    AO  &           109.9 & 0.5 & 25.9 & 0.1 & \phantom{$-$}35.0 & 0.1 & 2.8 \\
 &                  GBT &           117.1 & 0.1 & 25.7 & 0.1 & \phantom{$-$}34.6 & 0.1 & 2.3 \\
G039.883$-$0.346 &  AO  & \phantom{0}98.0 & 0.4 & 33.3 & 0.2 & \phantom{$-$}58.8 & 0.1 & 1.8 \\
 &                  GBT & \phantom{0}62.2 & 0.1 & 32.2 & 0.1 & \phantom{$-$}58.9 & 0.1 & 2.7 \\
G052.160+0.706 &    AO  & \phantom{0}13.3 & 0.5 & 29.0 & 1.4 & \phantom{$-$1}8.2 & 0.5 & 1.8 \\
 &                  GBT & \phantom{0}14.0 & 0.4 & 25.7 & 0.4 & \phantom{$-$1}7.9 & 0.4 & 2.8 \\
G061.720+0.864 &    AO  & \phantom{0}14.1 & 0.4 & 33.8 & 1.0 &           $-$73.4 & 0.4 & 1.9 \\
 &                  GBT & \phantom{0}18.6 & 0.1 & 27.6 & 0.1 &           $-$75.2 & 0.1 & 1.9 \\

\enddata
\tablenotetext{a}{AO = Arecibo Observatory; GBT = Green Bank Telescope}
\label{tab:comparison}
\end{deluxetable}
\clearpage


\begin{deluxetable}{lcccccccccl}
\tabletypesize{\scriptsize}
\tablecaption{Arecibo HRDS Kinematic Distances}
\tablewidth{0pt}
\tablehead{
\colhead{Name} &
\colhead{$V_{\rm lsr}$} & 
\colhead{$V_{\rm TP}$} & 
\colhead{$D_{\rm N}$} &
\colhead{$D_{\rm F}$} &
\colhead{N/F} & 
\colhead{$D_\sun$} &
\colhead{$\sigma_D$} &
\colhead{$R_{\rm gal}$} &
\colhead{$z$} &
\colhead{Notes\tablenotemark{a}} \\

\colhead{} &
\colhead{(\kms)} &
\colhead{(\kms)} &
\colhead{(kpc)} &
\colhead{(kpc)} &
\colhead{} &
\colhead{(kpc)} &
\colhead{(kpc)} &
\colhead{(kpc)} &
\colhead{(pc)} &
\colhead{}
}
\startdata

G034.591+0.244    &  $-19.4$  &  93.5  &  \nodata  &  16.1  &  F  &  16.1  &  1.1  &  10.3  &  \phn \phs 68.5  & Outer \\
G036.996$-$0.231  &  \phs 79.7  &  86.5  &  5.5  &  \phn 8.1  &  T  &  \phn 6.8  &  \nodata  &  \phn 5.3  &  \phn $-27.4$  & TP \\
G037.468$-$0.105  &  \phs 58.7  & 85.2 & 3.9     &  \phn 9.6  &  F  &  \phn 9.6  &  0.5 & \phn 5.9 & \phn $-17.6$  & W47 Complex \\
G039.195+0.226    &  $-21.6$  &  80.3  &  \nodata  &  15.4  &  F  &  15.4  &  1.1  &  10.3  &  \phn \phs 60.5  & Outer \\
G041.132$-$0.558  &  \phs 65.5  &  74.9  &  4.7  &  \phn 8.1  &  T  &  \phn 6.4  &  \nodata  &  \phn 5.9  &  \phn $-62.3$  & TP \\
G041.881+0.493    &  \phs 16.1  &  72.9  &  1.1  &  11.6  &  F  &  \phn 1.1  &  0.1  &  \phn 7.7  &  \phn \phn \phs 9.1  & HI\,E/A \\
G043.738+0.114    &  \phs 73.1  &  68.0  &  6.1  &  \phn 6.1  &  T  &  \phn 6.1  &  \nodata  &  \phn 5.9  &  \phn \phs 12.2  & TP \\
G045.039$-$0.643  &  \phs 61.3  &  64.6  &  4.9  &  \phn 7.1  &  T  &  \phn 6.0  &  \nodata  &  \phn 6.1  &  \phn $-67.3$  & TP \\
G046.176+0.536    &  \phn \phs 6.3  &  61.7  &  0.3  &  11.5  &  F  &  11.5  &  0.7  &  \phn 8.3  &  \phs 107.2  & G46 Complex \\
G046.203+0.535    &  \phn \phs 2.1  &  61.6  &  \nodata  &  11.8  &  F  &  11.8  &  0.8  &  \phn 8.5  &  \phs 109.9  & G46 Complex \\
G046.213+0.548    &  \phn \phs 5.8  &  61.6  &  0.3  &  11.5  &  F  &  11.5  &  0.8  &  \phn 8.3  &  \phs 109.8  & G46 Complex \\
G047.094+0.492    &  $-54.5$  &  59.4  &  \nodata  &  17.5  &  F  &  17.5  &  1.4  &  13.2  &  \phs 150.2  & Outer \\
G047.100+0.480    &  $-56.9$  &  59.4  &  \nodata  &  17.8  &  F  &  17.8  &  1.5  &  13.6  &  \phs 149.4  & Outer \\
G049.738$-$0.616  &  \phs 62.4  &  53.0  &  5.5  &  \phn 5.5  &  T  &  \phn 5.5  &  \nodata  &  \phn 6.5  &  \phn $-59.1$  & TP \\
G050.039$-$0.274  &  \phs 60.9  &  52.3  &  5.5  &  \phn 5.5  &  T  &  \phn 5.5  &  \nodata  &  \phn 6.5  &  \phn $-26.1$  & TP \\
G050.489+0.993    &  \phs 57.1  &  51.2  &  5.4  &  \phn 5.4  &  T  &  \phn 5.4  &  \nodata  &  \phn 6.6  &  \phn \phs 93.7  & TP \\
G050.785+0.168    &  \phs 47.3  &  50.6  &  4.2  &  \phn 6.5  &  T  &  \phn 5.4  &  \nodata  &  \phn 6.7  &  \phn \phs 15.7  & TP \\
G051.402$-$0.890  &  \phs 72.8  &  49.1  &  5.3  &  \phn 5.3  &  T  &  \phn 5.3  &  \nodata  &  \phn 6.6  &  \phn $-82.4$  & TP \\
G052.201+0.752    &  \phn \phs 8.7  &  47.3  &  0.5  &  \phn 9.9  &  F  &  \phn 9.9  &  0.8  &  \phn 8.2  &  \phs 130.1  & HI\,E/A; G52 \\
G052.259+0.700    &  \phn \phs 7.4  &  47.2  &  0.4  &  10.0  &  F  &  10.0  &  0.8  &  \phn 8.3  &  \phs 122.2  & HI\,E/A; G52 \\
G052.397$-$0.580  &  \phs 67.2  &  46.9  &  5.2  &  \phn 5.2  &  T  &  \phn 5.2  &  \nodata  &  \phn 6.7  &  \phn $-52.5$  & TP \\
G055.159$-$0.298  &  \phs 33.7  &  40.9  &  3.0  &  \phn 6.7  &  T  &  \phn 4.9  &  \nodata  &  \phn 7.2  &  \phn $-25.2$  & TP \\
G063.137+0.252    &  \phs 17.1  &  25.9  &  1.6  &  \phn 6.1  &  T  &  \phn 3.8  &  \nodata  &  \phn 7.9  &  \phn \phs 16.9  & TP \\

\enddata
\tablenotetext{a}{``TP'' = Tangent point distance; ``Outer'' = Outer Galaxy}
\label{tab:dist}
\end{deluxetable}
\clearpage

\end{document}